\newcommand{\home}{n:}          

\newcommand{\bw}{\mathbf{w}}

\newcommand{\bone}{\mathbf{1}}
\newcommand{\bzero}{\mathbf{0}}

\documentclass[a4paper,12pt,titlepage]{article}

\newcommand {\cint}      {confidence interval}

\newcommand {\nfci}      {95\% \cint}

\newcommand {\E}      [1] {{\mbox{E}}\left[#1\right]}

\newcommand {\probability}   [1] {{\rm{P}}\left(#1\right)}

\newcommand {\varb}   [1] {{\rm var}\left(#1\right)}

\newcommand {\cov}    [1] {{\rm cov}\left(#1\right)}

\newcommand {\varhat} [1] {\widehat{\rm var}\left(#1\right)}

\newcommand {\logit}  [1] {{\rm logit}\ #1}

\newcommand {\ben}        {\begin{enumerate}}
\newcommand {\een}        {\end{enumerate}}
\newcommand {\bi}        {\begin{itemize}}
\newcommand {\ei}        {\end{itemize}}
\newcommand {\beq}        {\begin{equation}}

\newcommand {\beqa}        {\begin{eqnarray}}
\newcommand {\beqas}        {\begin{eqnarray*}}
\newcommand {\eeq}        {\end{equation}}

\newcommand {\eeqa}        {\end{eqnarray}}
\newcommand {\eeqas}        {\end{eqnarray*}}
\newcommand {\bt} [1]    {\begin{center} \begin{tabular}{#1} }
\newcommand {\btc} [2]   {\begin{center} #1 \nopagebreak \vspace{1ex}\\ \begin{tabular}{#2} }
\newcommand {\et}        {\end{tabular} \end{center}}
\newcommand {\file}[1]{}

\newcommand {\type}[1]   {\verb|#1|}

\usepackage{epsfig}
\usepackage{citesort}
\usepackage{pdfsync}
\usepackage{natbib}
\usepackage{pdflscape}
\usepackage{xcolor}

\renewcommand{\baselinestretch}{2}

\title{A mean score method for sensitivity analysis to departures from the missing at random assumption in randomised trials}
\author{
Ian R. White$^{1,2,*}$,
James Carpenter $^{2,3}$, and
Nicholas J. Horton $^4$\\
$^1$ MRC Biostatistics Unit, Cambridge, UK\\
$^2$ MRC Clinical Trials Unit at UCL, London, UK\\
$^3$ London School of Hygiene and Tropical Medicine, UK\\
$^4$ Amherst College, Amherst, MA, USA\\
$^*$ Email: ian.white@ucl.ac.uk
}

\usepackage[pdfborder={0 0 0},
pdfauthor={Ian White},
pdftitle={A mean score method for sensitivity analysis to departures from the missing at random assumption in randomised trials}]{hyperref}

\begin{document}


\maketitle

\begin{abstract}
Most analyses of randomised trials with incomplete outcomes make untestable assumptions and should therefore be subjected to sensitivity analyses. However, methods for sensitivity analyses are not widely used. We propose a mean score approach for exploring global sensitivity to departures from missing at random or other assumptions about incomplete outcome data in a randomised trial. We assume a single outcome analysed under a generalised linear model. One or more sensitivity parameters, specified by the user, measure the degree of departure from missing at random in a pattern mixture model. Advantages of our method are that its sensitivity parameters are relatively easy to interpret and so can be elicited from subject matter experts; it is fast and non-stochastic; and its point estimate, standard error and confidence interval agree perfectly with standard methods when particular values of the sensitivity parameters make those standard methods appropriate.
We illustrate the method using data from a mental health trial.


\emph{Running title}:
Mean score method for sensitivity analysis.
\end{abstract}


\section{Introduction}

Missing outcome data are a threat to the validity of randomised controlled trials, and they usually require untestable assumptions to be made in the analysis.
One common assumption is that data are missing at random (MAR) \citep{LittleRubin02}.  Other possible assumptions may be less implausible in particular clinical settings. For example, in smoking cessation trials, the outcome is binary, indicating whether an individual quit over a given period, and it is common to assume that missing values are failures --- ``missing=failure'' \citep{West++05}; while in weight loss trials, missing data is sometimes assumed to be unchanged since baseline --- ``baseline observation carried forward'' \citep{Ware03}.

The US \citet{CNSTAT10} suggested measures that should be taken to minimise the amount of missing outcome data in randomised trials, and described analysis strategies based on various assumptions about the missing data.
This report recommended that ``Sensitivity analyses should be part of the primary reporting of findings from clinical trials. Examining sensitivity to the assumptions about the missing data mechanism should be a mandatory component of reporting.'' However, among ``several important areas where progress is particularly needed'', the first was ``methods for sensitivity analysis and principled decision making based on the results from sensitivity analyses''.
Sensitivity analysis is also an essential part of an intention-to-treat analysis strategy, which includes all randomised individuals in the analysis strategy \citep{ian:ITTmed,ian:ITTstat}: even if the main analysis is performed under MAR and hence draws no information from individuals with no outcome data, such individuals are included in sensitivity analysis and hence in the analysis strategy.


Sensitivity analysis is often done by performing two different analyses, such as an analysis assuming MAR and an analysis by last observation carried forward, and concluding that inference is robust if the results are similar \citep{ian:missing_survey}. Better is a principled sensitivity analysis, where the data analyst typically formulates a model including unidentified `sensitivity parameter(s)' that govern the degree of departure from the main assumption (e.g.\ from MAR), and explores how the estimate of interest varies as the sensitivity parameter(s) are varied \citep{Rotnitzky++98,Kenward++01}. We consider global sensitivity analyses where the sensitivity parameter(s) are varied over a range of numerical values that subject-matter experts consider plausible.

Likelihood-based analyses assuming MAR can usually ignore the missing data mechanism and simply analyse the observed data \citep{LittleRubin02}. Under a missing not at random (MNAR) assumption, however, it is usually necessary to model the data of interest jointly with the assumed missing data mechanism.
The joint model can be specified as
a pattern-mixture model, which explicitly describes the differences between profiles of patients who complete and drop out \citep{Little93,Little94},
or as a selection model, which relates the chance of drop-out to the (possibly missing) response values either directly \citep{DiggleKenward94,Kenward98} or indirectly through a random effect \citep{FollmannWu95,Roy03}.
\citet{Rotnitzky++98} proposed a selection model for incomplete repeated measures data and showed how to estimate it by inverse probability weighting, given values of the  sensitivity parameters.
\citet{Scharfstein++03a} adopted a non-parametric Bayesian approach to analysing incomplete randomised trial data, and argued that sensitivity parameters are more plausibly a priori independent of other parameters of interest in a selection model than in a pattern-mixture model.
\citet{Scharfstein++14} proposed a fully parametric approach based on a selection model.
On the other hand, \citet{DanielsHogan00} advocated a pattern-mixture framework as ``a convenient and intuitive framework for conducting sensitivity analyses''.

We use the pattern-mixture model in this paper because its sensitivity parameters are usually more easily interpreted \citep{ian:IMpriors,ian:metamiss1}.
For a binary outcome, a convenient sensitivity parameter is the informative missing odds ratio (IMOR),
defined, conditional on covariates, as the odds of positive outcome in missing values divided by the odds of positive outcome in observed values
\citep{ian:metamiss0,Kaciroti++09}. For a continuous outcome, a convenient sensitivity parameter is the covariate-adjusted mean difference between missing and observed outcomes \citep{ian:metamisscts}.

Most estimation procedures described for general pattern-mixture models are likelihood-based \citep{Little93,Little94,LittleYau96,HedekerGibbons06}, while the \citet{CNSTAT10} describes point estimation using sensitivity parameters with bootstrap standard errors. In this paper we propose instead using the mean score method, a computationally convenient method which was originally proposed under a MAR assumption for incomplete outcome data \citep{Pepe++94} and for incomplete covariates \citep{ReillyPepe95}. The method is particularly useful to allow for auxiliary variables, so that outcomes can be assumed MAR given model covariates and auxiliary variables but not necessarily MAR given model covariates alone \citep{Pepe++94}. We are not aware of the mean score method having been used for sensitivity analysis.

The aim of this paper is to propose methods for principled sensitivity analysis that are fast, non-stochastic, available in statistical software, and agree exactly with standard methods in the special cases where standard methods are appropriate.
We focus on randomised trials with outcome measured at a single time, allowing for continuous or binary outcomes, or indeed any generalised linear model, and for covariate adjustment.

The paper is organised as follows.
Section \ref{sec:meanscore} describes our proposed method.
Section \ref{sec:equiv} proposes small-sample corrections which yield exact equivalence to standard procedures in special cases.
Section \ref{sec:QUATRO} illustrates our method in QUATRO, a mental health trial with outcome measured at a single time.
Section \ref{sec:sim} describes a simulation study.
Section \ref{sec:discuss} discusses the implementation of our method, possible alternatives, limitations and extensions.


\section{Mean score approach} \label{sec:meanscore}

\newcommand{\bx}{\mathbf{x}}
\newcommand{\bz}{\mathbf{z}}
\newcommand{\bbeta}{\mbox{\boldmath $\beta$}}
\newcommand{\balpha}{\mbox{\boldmath $\alpha$}}
\newcommand{\bU}{\mathbf{U}}
\newcommand{\bB}{\mathbf{B}}
\newcommand{\bC}{\mathbf{C}}
\newcommand{\bV}{\mathbf{V}}
\newcommand{\Beta}{B}

Assume that for the $i$th individual ($i = 1$ to $n$) in an individually randomised trial, there is an outcome variable $y_i$, and let $r_i$ be an indicator of $y_i$ being observed.
Let $n_{obs}$ and $n_{mis}=n-n_{obs}$ be the numbers of observed and missing values of $y$ respectively.
Let $\bx_{i}$ be a vector of covariates including
the $p_S$-dimensional fully-observed covariates $\bx_{Si}$ in the substantive model, comprising an intercept, an indicator $z_i$ for the randomised group, and (optionally) baseline covariates.
$\bx_{i}$ may also include
fully-observed auxiliary covariates $\bx_{Ai}$ that are not in the substantive model but that help to predict $y_i$, and/or
covariates $\bx_{Ri}$ that are only observed in individuals with missing $y_i$ and describe the nature of the missing data: for example, the reason for missingness.

The aim of the analysis is to estimate the effect of randomised group, adjusting for the baseline covariates. We assume the substantive model is a generalised linear model (GLM) with canonical link,
\begin{equation}\label{eq:SM}
\E{y_i|\bx_{Si};\bbeta_S} = h(\bbeta_S^T \bx_{Si})
\end{equation}
where $h(.)$ is the inverse link function. We are interested in estimating $\beta_{Sz}$, the component of the $p_S$-dimensional vector $\bbeta_S$ corresponding to $z$.

If we had complete data, we would estimate $\bbeta_S$ by solving the estimating equation $U_S^*(\bbeta_S)=0$ where $U_S^*(\bbeta_S)=\sum_i U_{Si}^*(\bbeta_S)$ and
\begin{equation}\label{eq:score:complete}
U_{Si}^*(\bbeta_S) = \{y_i - h(\bbeta_S^T \bx_{Si})\} \bx_{Si}.
\end{equation}

The mean score approach \citep{Pepe++94,ReillyPepe95} handles missing data by replacing $U_{S}^*(\bbeta_S)$ with $U_{S}(\bbeta_S)$, its expectation over the distribution of the missing data given the observed data.
We write $U_S(\bbeta_S)=\sum_i U_{Si}(\bbeta_S)$
and $U_{Si}(\bbeta_S) = \E{ U_{Si}^*(\bbeta_S) | \bx_{i}, r_i, r_i y_i }$.
Then by the repeated expectation rule, $\E{U_{Si}(\bbeta_S) | \bx_{Si}} = \E{U_{Si}^*(\bbeta_S) | \bx_{Si}}$ since $\bx_{i}$ includes $\bx_{Si}$, so $U_S(\bbeta_S)=0$ is an unbiased estimating equation if $U_S^*(\bbeta_S)=0$ is.

To compute $U_{Si}(\bbeta_S)$, we need only $\E{y_i | \bx_{i}, r_i=0}$, because (\ref{eq:score:complete}) is linear in  $y_i$.
We estimate this using the pattern-mixture model
\begin{equation}\label{eq:PM}
\E{y_i|\bx_{i},r_i;\bbeta_P} = h\left(\bbeta_P^T \bx_{Pi} + \Delta(\bx_{i}) (1-r_i)\right)
\end{equation}
where $\bx_{Pi}=(\bx_{Si},\bx_{Ai})$ of dimension $p_P$;
the subscript $P$ distinguishes the parameters $\bbeta_P$ and covariates $\bx_{Pi}$ of the pattern-mixture model from the parameters $\bbeta_S$ and covariates $\bx_{Si}$ of the substantive model.
Models (\ref{eq:SM}) and (\ref{eq:PM}) are typically not both correctly specified: we return to this issue in the simulation study.

In equation (\ref{eq:PM}), $\Delta(\bx_{i})$ is a user-specified departure from MAR for individual $i$.
MAR in this setting means that $p(r_i=1|\bx_i,y_i)=p(r_i=1|\bx_i)$, which implies $\E{y_i|\bx_i,r_i}=\E{y_i|\bx_i}$ and hence $\Delta(\bx_{i})=0$ for all $i$.
A simple choice of $\Delta(\bx_{i})$ that expresses MNAR is $\Delta(\bx_{i}) = \delta$, where the departure from MAR is the same for all individuals.
Differences in departure from MAR between randomised groups are often plausible and can have strong impact on the estimated treatment effect \citep{ian:IMpriors}, so an alternative choice is $\Delta(\bx_{i}) = \delta_{z_i}$.
The departure $\Delta(\bx_{i})$ could also depend on reasons for missingness coded in $\bx_{Ri}$: for example, it could be 0 for individuals lost to follow-up (if MAR seemed plausible for them), and $\delta_{z_i}$ for individuals who refused follow-up.

Putting it all together, the mean score method solves
\begin{equation}\label{eq:score:inc}
\sum_i \{\tilde{y}_i(\bbeta_P) - h(\bbeta_S^T \bx_{Si})\} \bx_{Si} = \bzero
\end{equation}
where $\tilde{y}_i(\bbeta_P)$ is defined as $y_i$ if $r_i=1$ and $h\left(\bbeta_P^T \bx_{Pi} + \Delta(\bx_{i}) (1-r_i)\right)$ if $r_i=0$.

\subsection{Estimation using full sandwich variance} \label{sec:fullsandwich}

The parameter $\bbeta_P$ in (\ref{eq:PM}) is estimated by regressing $y_i$ on $\bx_{Pi}$ in the complete cases ($r_i=1$).
Once $\bbeta_P$ is estimated, we calculate the $\tilde{y}_i(\hat{\bbeta}_P)$ using the known values $\Delta(\bx_{i})$ and solve (\ref{eq:score:inc}) for $\bbeta_S$.
The whole procedure amounts to solving the set of estimating equations $\bU(\bbeta)=\bzero$ where
      $\bbeta=(\bbeta_S^T,\bbeta_P^T)^T$,
      $\bU(\bbeta)=(\bU_{S}(\bbeta)^T,\bU_{P}(\bbeta_P)^T)^T$,
      $\bU_S(\bbeta)=\sum_i \bU_{Si}(\bbeta)$,
      $\bU_P(\bbeta_P)=\sum_i \bU_{Pi}(\bbeta_P)$,
\begin{equation}\label{eq:score:bi}
\left.
\begin{array}{rcl}\bU_{Si}(\bbeta)&=&\left\{r_i y_i + (1-r_i) h(\bbeta_P^T \bx_{Pi} + \Delta(\bx_{i})) - h(\bbeta_S^T \bx_{Si})\right\}\bx_{Si}, \\
\bU_{Pi}(\bbeta_P)&=&r_i \left\{y_i - h(\bbeta_P^T \bx_{Pi})\right\}\bx_{Pi} .\end{array}
\right\}
\end{equation}

\citet{Pepe++94} derived a variance expression for $\bbeta_S$ assuming categorical $\bx_{Si}$. To accommodate any form of $\bx_{Si}$, we instead obtain standard errors by the sandwich method, based on both estimating equations.
The sandwich estimator of $\varb{\hat{\bbeta}}$ is
\begin{equation} \label{eq:sandwich}
\bV=\bB^{-1}\bC \bB^{-T}
\end{equation}
where
$\bB=-d\bU/d\bbeta$ evaluated at $\bbeta=\hat{\bbeta}$,
$\bC=\sum_i \bU_i(\hat{\bbeta}) \bU_i(\hat{\bbeta})^T$
and
$\bU_i(\bbeta)=(\bU_{Si}(\bbeta)^T,\bU_{Pi}(\bbeta_P)^T)^T$.
$\bB$ and $\bC$ are given in Section \ref{app:variance} of the Supplementary Materials.

\subsection{Estimation using two linear regressions} \label{sec:twolinreg}

A special case arises if there are no auxiliary variables, so $\bx_{Pi}=\bx_{Si}$ for all $i$, and $h(.)$ is the identity function, as in linear regression. Then we can rearrange (\ref{eq:score:bi}) to give
$$
\bU_{Si}(\bbeta) - \bU_{Pi}(\bbeta_P)
=
\left\{
(1-r_i) \Delta(\bx_{i})
-
(\bbeta_S - \bbeta_P)^T\bx_{Si}\right\}\bx_{Si}.
$$
Thus $(\bbeta_S - \bbeta_P)$ may be estimated by linear regression of $(1-r_i) \Delta(\bx_{i})$ on $\bx_{Si}$.
This estimate is uncorrelated with $\hat{\bbeta}_P$ because
$\cov{\bU_{Si}(\bbeta) - \bU_{Pi}(\bbeta_P), \bU_{Pi}(\bbeta_P)}=\bzero$. This gives a direct way to estimate $\bbeta_S$, and its variance $\varb{\hat{\bbeta}_P} + \varb{\hat{\bbeta}_S-\hat{\bbeta}_P}$, from standard linear regressions.

In particular, consider a two-arm trial with no covariates, and write the coefficients of $z_i$ in (\ref{eq:SM}) and (\ref{eq:PM}) as $\beta_{Sz}$ and $\beta_{Pz}$. Then  $\beta_{Sz} - \beta_{Pz}$ is estimated as the difference between arms in the mean of $(1-r_i) \Delta(\bx_{i})$, which is $a_1 \delta_1 - a_0 \delta_0$ where $a_j$ is the proportion of missing data in arm $j=0,1$ and $\delta_j$ is the average of $\Delta(\bx_{i})$ over individuals with missing data in arm $j=0,1$. Therefore the estimated parameter of interest is
$\hat{\bbeta}_{Sz} = \hat{\bbeta}_{Pz} + a_1 \delta_1 - a_0 \delta_0$
as in \citet{ian:IMpriors}; the same result can be derived in other ways.


\section{Equivalence to standard procedures} \label{sec:equiv}

We now consider two special cases which can be fitted by standard procedures:
(1) when MAR is assumed and there are no auxiliary variables, so incomplete cases contribute no information and the standard procedure is an analysis of complete cases,
and (2) when `missing = failure' is assumed for a binary outcome, so the standard procedure is to replace missing values with failures.
Our aim is that point estimates, standard errors and confidence intervals produced by the mean score procedure should agree exactly with those produced by the standard procedures in these cases.

Equality of point estimates is easy to see. In case (1), we have $\Delta(\bx_{i})=0$ and $\bx_{Pi}=\bx_{Si}$ for all $i$, so if $\bbeta_P$ solves $\bU_{P}(\bbeta_P)=\bzero$ then $\bbeta=(\bbeta_P,\bbeta_P)$ solves $\bU_{S}(\bbeta)=\bzero$.
In case (2), `missing = failure' can be expressed as $\Delta(\bx_{i})=-\infty$ for all $i$, so the mean score procedure gives $y_i^*=0$ whenever $r_i=0$, and solving $\bU_{S}(\bbeta_S,\bbeta_P)=\bzero$ gives the same point estimate as replacing missing values with failures.

\newcommand{\neff}{n_{\mbox{\it\scriptsize eff}}}
\newcommand{\pstar}{p^*}

Exact equality of variances between mean score and standard procedures depends on which finite sample corrections (if any) are applied.
Many such corrections have been proposed to reduce the small-sample bias of the sandwich variance estimator and to improve \cint\ coverage \citep{KauermannCarroll01,Lu++07}.
Here, we assume that the standard procedures use the commonly used
small-sample correction factor for the sandwich variance $\varhat{\hat{\bbeta}_S} = f \bV$ where $f= n/(n-\pstar)$,
$n$ is the sample size and $\pstar$ is the number of regression parameters (in linear regression) or 1 (in other GLMs): this is for example the default in Stata \citep{Stata12}.

Exact equality of \cint s between mean score and standard procedures additionally depends on the distributional assumptions used to construct \cint s.
Here, we assume that standard procedures for linear regression construct confidence intervals from the $t$ distribution with $n-\pstar$ degrees of freedom, and that standard procedures for other GLMs construct confidence intervals from the Normal distribution.

With missing data, we propose using the same small-sample correction factor, distributional assumptions and degrees of freedom, but replacing $n$ by an effective sample size $\neff$ as shown below. Thus we propose forming \cint s for linear regression by assuming
$\hat{\bbeta}_S \sim t_{\neff-\pstar} \left(\bbeta_S,\frac{\neff}{\neff-\pstar}\bV\right)$
and for other GLMs by assuming
$\hat{\bbeta}_S \sim N\left(\bbeta_S,\frac{\neff}{\neff-\pstar}\bV\right)$.

\subsection{Full sandwich method}\label{sec:fullsandwich:small}

For the full sandwich method of Section \ref{sec:fullsandwich}, we propose computing $\neff$ as  $n_{obs} + (I_{mis}/I_{mis^*}) n_{mis}$ where $I_{mis}$ is the influence of the individuals with missing values, and $I_{mis^*}$ is the same individuals' influence if the missing values had been observed.
The comparison of the same individuals is crucial, because missing individuals, if observed, would have different influence from observed individuals.

To determine $I_{mis}$, we consider weighted estimating equations $\sum_i w_i \bU_i(\bbeta)=\bzero$ with solution $\bbeta_\bw$. Differentiating with respect to $\bw=(w_1,\ldots,w_n)^T$ at $\bw=\bone$ gives $\bU^T + \frac{d\bU(\bbeta)}{d\bbeta} \frac{d\bbeta_\bw}{d\bw} = \bzero$ where $\bU$ is a $n \times (p_S+p_P)$ matrix with $i$th row $\bU_i(\bbeta)$. Since also $\frac{d\bU(\bbeta)}{d\bbeta}=-\bB$, we get
\begin{equation}\label{eq:dbetadw}
\frac{d\bbeta_\bw}{d\bw}=-\bB^{-1}\bU^T.
\end{equation}
We now define the influence of observation $i$ as $$I_{mis,i} = \frac{d\bbeta_{\bw S}}{dw_i}^T \bV_S^{-1}\frac{d\bbeta_{\bw S}}{dw_i}$$ where the $S$ subscript denotes the elements corresponding to $\bbeta_S$. Hence we define the influence of the individuals with missing values as $I_{mis} = \sum_i (1-R_i) I_{mis,i}$.

To determine $I_{mis^*}$, we let $\bbeta_{\bw S}^*$ be the (unknown) parameter estimate that would be obtained if the complete data had been observed, following the pattern-mixture model (\ref{eq:PM}). In this case the influence would be
$$I_{mis,i}^* = \frac{d\bbeta_{\bw S}^*}{dw_i}^T \bV_S^{-1}\frac{d\bbeta_{\bw S}^*}{dw_i}.$$
We define the ``full-data influence'' as the expectation of $I_{mis,i}^*$ over the distribution of the complete data given the observed data, under the pattern-mixture model (\ref{eq:PM}).
From (\ref{eq:dbetadw}) we get $\frac{d\bbeta_\bw^*}{dw_i}=-\bB^{-1}\bU^{*T}_i$ and $\bU^{*T}_i=\left\{y_i^*-h(\bbeta_s^T \bx_{Si}) \right\} \bx_{Si}$, so that
\begin{equation}\label{eq:Istar}
  \E{I_{mis,i}^*} = \E{\left\{y_i^*-h(\bbeta_s^T \bx_{Si}) \right\}^2} \bx_{Si}^T \bB_{SS}^{-T} \bV_S^{-1}\bB_{SS}^{-1}\bx_{Si}
\end{equation}
and $\E{\left\{y_i^*-h(\bbeta_s^T \bx_S) \right\}^2}$ is evaluated as the squared residual plus the residual variance from model (\ref{eq:SM}) for individual $i$.
Finally, $I_{mis^*} = \sum_i (1-r_i) \E{I_{mis,i}^*}$.
In case (1), $I_{mis}=0$ so $\neff=n_{obs}$, as in standard analysis.
In case (2), $I_{mis^*}=I_{mis}$ so $\neff=n$, again as in standard analysis.

\subsection{Two linear regressions method}\label{sec:twolinreg:small}

For the two linear regressions method of Section \ref{sec:twolinreg}, the small-sample correction to the variance is naturally applied separately to each variance  in
$\varhat{\hat{\bbeta}_P}+ \varhat{\hat{\bbeta}_S-\hat{\bbeta}_P}$ $= V_{small}$ say.
We can derive the corresponding variances without small-sample correction as
$\frac{n_{obs}-p}{n_{obs}}\varhat{\hat{\bbeta}_P}+ \frac{n-p}{n}\varhat{\hat{\bbeta}_S-\hat{\bbeta}_P}=V_{large}$ say.
To compute $\neff$, we use the heuristic $V_{small} \approx \frac{\neff}{\neff-p} V_{large}$, and hence we estimate $\neff$ by solving
$|V_{small}| = \left(\frac{\neff}{\neff-p}\right)^{p} |V_{large}|$.
In case (1), $\varhat{\hat{\bbeta}_S-\hat{\bbeta}_P}=\bzero$ so $\neff=n_{obs}$, as in standard analysis. Case (2) does not apply to linear regression.


\section{Example: QUATRO trial} \label{sec:QUATRO}

The QUATRO trial \citep{Gray++06} was a randomised controlled trial in people with schizophrenia, to evaluate the effectiveness of a patient-centred intervention to improved adherence to prescribed antipsychotic medications. The trial included 409 participants in four European centres. The primary outcome, measured at baseline and 1 year, was participants' quality of life, expressed as the mental health component score (MCS) of the SF-36 \citep{Ware93}. The MCS is designed to have mean 50 and standard deviation 10 in a standard population, and a higher MCS score implies a better quality of life.
The data are summarised in Table \ref{tab:data}.

We first estimate the intervention effect on MCS, adjusted for baseline MCS and centre. Thus in the substantive model (\ref{eq:SM}), $h(.)$ is the identity link, $y_i$ is MCS at 1 year for participant $i$, and $\bx_{Si}$ is a vector containing 1, randomised group $z_i$ (1 for the intervention group and 0 for the control group), baseline MCS and dummy variables for three centres. We have no $\bx_{Ai}$ or $\bx_{Ri}$.

We also estimate the intervention effect on a binary variable, MCS dichotomised at an arbitrary value of 40, where for illustration we use baseline MCS and dummy variables for centre as auxiliary variables $\bx_{Ai}$. Thus in the substantive model (\ref{eq:SM}), $h(\eta)=1/\{1+\exp{(-\eta)}\}$ is the inverse of the logit link, $y_i$ is dichotomised MCS at 1 year for participant $i$, and $\bx_{Si}$ is a vector containing 1 and $z_i$.

In both analyses, we replace the 23 missing values of baseline MCS with the mean baseline MCS: while such mean imputation is not valid in general, it is appropriate and efficient in the specific case of estimating intervention effects with missing baseline covariates in randomised trials \citep{ian:MSGBSL,ian:MissInd}.

As expected (results not shown),
the point estimate, standard error and confidence interval from the mean score method agree exactly with standard methods under MAR and under missing=failure, using the small-sample corrections of Section \ref{sec:equiv}.

We consider three sets of sensitivity analyses using the mean score method around a MAR assumption, with departures from MAR (1) in the intervention arm only ($\Delta(\bx_{i})=\delta z_i$), (2) in both arms ($\Delta(\bx_{i})=\delta$), or (3) in the control arm only ($\Delta(\bx_{i})=\delta(1-z_i)$) \citep{ian:ITTstat}.
For the quantitative outcome, the investigators suggested that the mean of the missing data could plausibly be lower than the mean of the observed data by up to 10 units (equal to nearly one standard deviation of the observed data), so we allow $\delta$ to range from 0 to -10 \citep{ian:QUATRO_MNAR}.
The investigators were not asked about missing values of the dichotomised outcome, so for illustrative purposes we allow $\delta$ to range from 0 to -6, which is close to ``missing=failure''.

\begin{figure}[t]
\caption{QUATRO trial: sensitivity analysis for the estimated intervention effect on the MCS (with \nfci) over a range of departures from MAR.}
\vspace{1ex}
\label{fig:QUATROsens}
\includegraphics[scale=0.8]{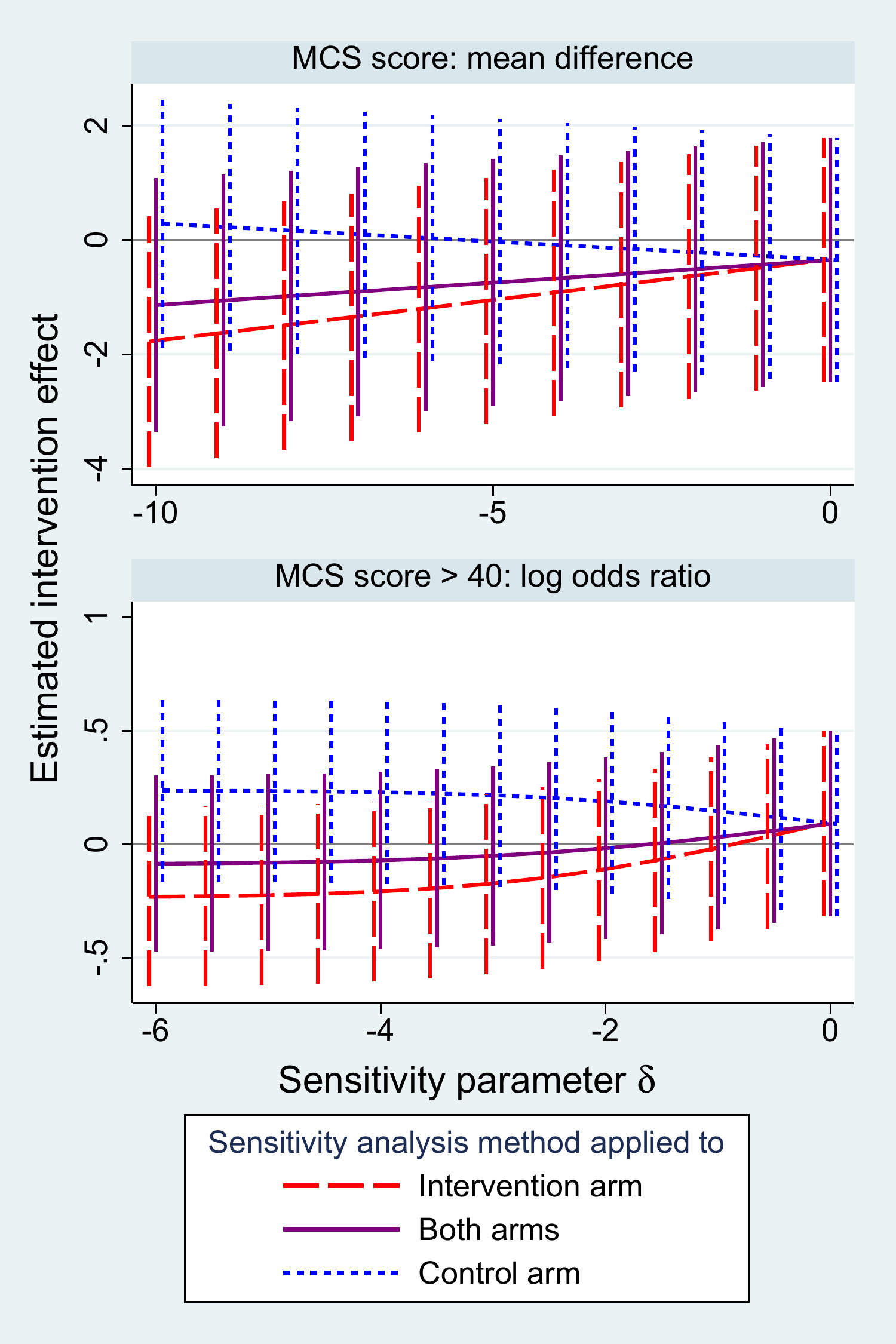}
\end{figure}

Figure \ref{fig:QUATROsens} shows the results of these sensitivity analysis using the two linear regressions method for the quantitative outcome (upper panel) and using the full sandwich method for the dichotomised outcome (lower panel). Results are more sensitive to departures from MAR in the intervention arm because there are more missing data in this arm. However, the finding of a non-significant intervention effect is unchanged over these ranges of sensitivity analyses.
This means that the main results of the trial are robust even to quite strong departures from MAR.

\begin{figure}[t]
\caption{QUATRO data: effective sample size in sensitivity analysis.}
\vspace{1ex}
\label{fig:QUATROneff}
\includegraphics[scale=0.8]{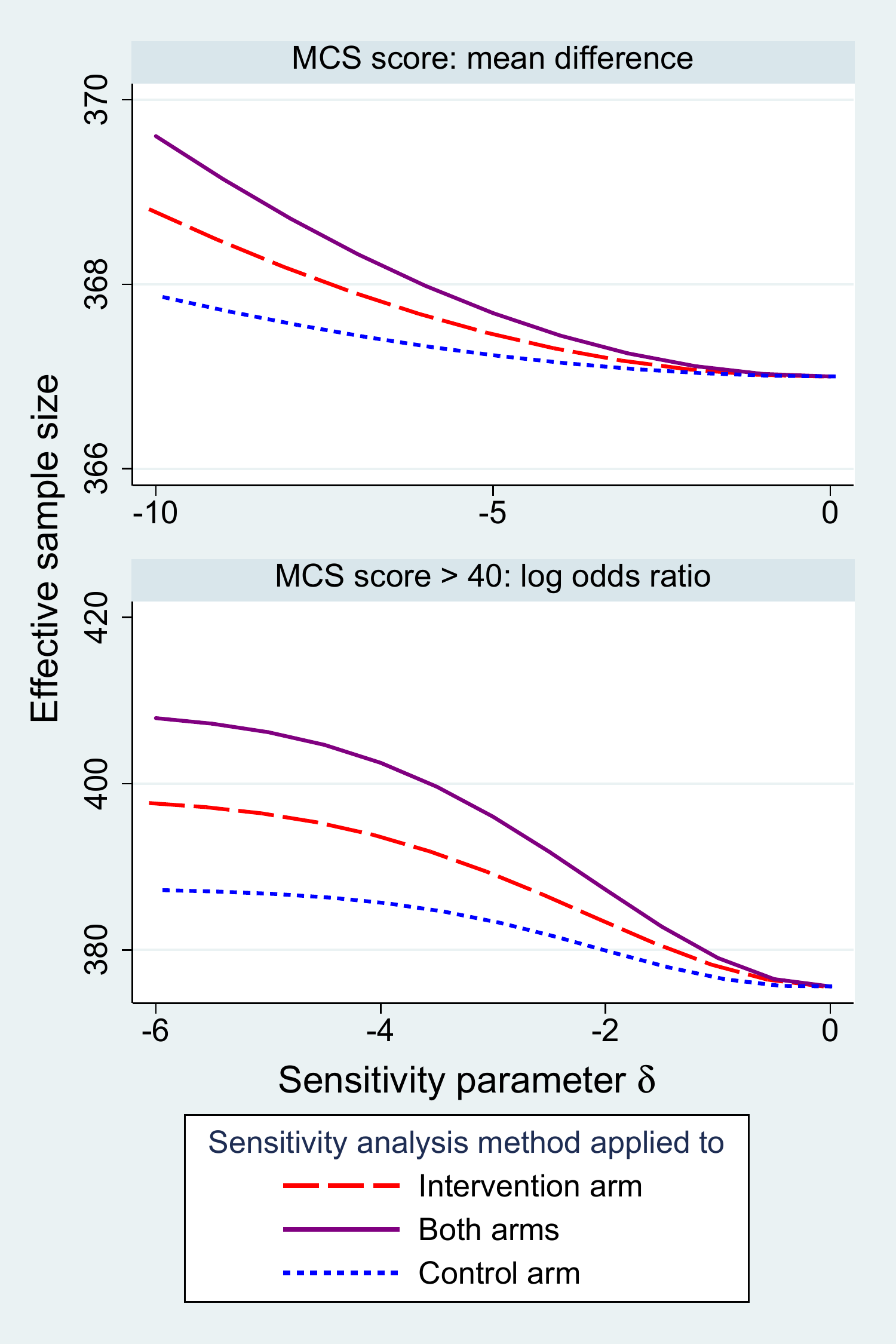}
\end{figure}

Figure \ref{fig:QUATROneff} shows the effective sample size for these two analyses. Effective sample size increases from 367 at MAR for both analyses. For the dichotomised outcome it is near the total sample size of 409 when $\delta=-6$ in both arms. For the quantitative outcome it does not pass 370 because the range of $\delta$ is more moderate.


\section{Simulation study} \label{sec:sim}

We report a simulation study aiming (i) to evaluate the performance of the mean score method when it is correctly specified, (ii) to compare the mean score method with alternatives, and (iii) to explore the impact of the incompatibility of models (\ref{eq:SM}) and (\ref{eq:PM}). We assume the sensitivity parameters $\Delta(\bx_{i})$ are correctly specified.

\subsection{Data generating models}

We generate data under four data generating models (DGMs), each with four choices of parameters. We focus on the case of a binary outcome.
In DGMs 1-3, we generate data under a pattern-mixture model.
In DGM 1, there are no baseline covariates, so $\bx_i=\bx_{Si}=\bx_{Pi}=(1,z_i)$.
We generate a treatment indicator $z_i \sim Bern(0.5)$;
a missingness indicator $r_i$ with $\logit{\probability{r_i|z_i}}=\alpha_1 + \alpha_z z_i$;
and a binary outcome $y_i$ following model (\ref{eq:PM}) with $\logit{\probability{y_i|z_i,r_i}}=\beta_{P1} + \beta_{Pz} z_i + \beta_{Pr} (1-r_i)$.
The substantive model (\ref{eq:SM}) is then
$\logit{\probability{y_i|z_i}}=\beta_{S1} + \beta_{Sz} z_i$.
Because this substantive model contains only a single binary covariate, it is saturated and cannot be mis-specified.
Therefore both substantive model and pattern-mixture model are correctly specified

DGM 2 extends DGM 1 by including  $x_i \sim N(0,1)$ as a single baseline covariate independent of $z_i$, so $\bx_i=(1,x_i,z_i)$.
The missingness indicator follows
$\logit{\probability{r_i|x_i,z_i}}=\alpha_1 + \alpha_x x_i + \alpha_z z_i$
and the binary outcome follows model (\ref{eq:PM}) with
$\logit{\probability{y_i|x_i,z_i,r_i}}=\beta_{P1} + \beta_{Px} x_i + \beta_{Pz} z_i + \beta_{Pr} (1-r_i)$.
The substantive model is as in DGM 1 with $\bx_{Si}=(1,z_i)$, and $\bx_{Ai}=(x_i)$ is an auxiliary variable in the analysis.
Thus the substantive model  and pattern-mixture model  are again both correctly specified.

DGM 3 is identical to DGM 2, but now $x_i$ is included in the substantive model, which is therefore
$\logit{\probability{y_i|x_i,z_i}}=\beta_{S1} + \beta_{Sx} x_i + \beta_{Sz} z_i$
with $\bx_{Si}=\bx_{Pi}=(1,x_i,z_i)$.
Now the substantive model is incorrectly specified while the pattern-mixture model  remains correctly specified.

DGM 4 is a selection model. Here $z_i$ and $x_i$ are generated as in DGM 2, then $y_i$ is generated following the substantive model
$\logit{\probability{y_i|x_i,z_i}}=\beta_{S1} + \beta_{Sx} x_i + \beta_{Sz} z_i$
and
$r_i$ is generated following
$\logit{\probability{r_i|x_i,z_i,y_i}}=\alpha_1 + \alpha_x x_i + \alpha_z z_i + \alpha_y y_i$.
Here the substantive model is correctly specified while the pattern mixture model is mis-specified.

For the parameter values, we consider scenarios a-d for each DGM.
In scenario a, the sample size is $n_{obs}=500$;
the missingness model has $\alpha_x=\alpha_z =\alpha_y =1$ and we choose
$\alpha_1$ to fix $\pi_{obs}=\probability{r=1}=0.75$;
and the pattern mixture model has
$\beta_{P1} =0$,
$\beta_{Px}=\beta_{Pz}  =1$,
$\beta_{Pr}  =-1$.
($\alpha_x$ and $\beta_{Px}$ are ignored in DGM 1, $\alpha_y$ is ignored in DGM 1-3 and $\beta_{Pr}$ is ignored in DGM 4.)
Scenarios b-d vary scenario a by setting
$n_{obs}=2000$,
$\pi_{obs}=0.5$,
and $\beta_{Pr}  =-2$ respectively.
1000 data sets were simulated in each case.
Table \ref{tab:simdesc} summarises the simulation design.

\subsection{Analysis methods}

The mean score (MS) method is implemented as described in sections \ref{sec:meanscore} and \ref{sec:equiv}, with logit link.
$x_i$ is used as an auxiliary in DGM 2.
In DGM 1-3, $\Delta(\bx_{i})$ is taken to equal the known value $-\beta_{Pr}$ for all individuals; in DGM 4, $\Delta(\bx_{i})$ is not known but (for the purposes of the simulation study) is estimated by fitting the pattern-mixture model to a data set of size 1,000,000 before data deletion.

The MS method is compared with analysis of data before data deletion (Full); analysis of complete cases (CC), which wrongly assumes MAR; and two alternative methods that allow for MNAR, multiple imputation (MI) and selection model with inverse probability weighting (SM).

In the MI approach \citep{Rubin87,ian:MItutorial}, the imputation model is equation (\ref{eq:PM}), and data are imputed with an offset $\Delta(\bx_{i})$ in the imputation model. The number of imputations is fixed at 30.

In the SM approach, we use the response model $\logit{p(r_i=1|y_i,\bx_{i})}=\balpha^T \bx_{Pi} + \Delta^*(\bx_{i}) y_i$
where the sensitivity parameter $\Delta^*(\bx_{i})$ expresses departure from MAR as the log odds ratio for response per 1-unit change in $y_i$.
In DGM 4, $\Delta^*(\bx_{i})$ is taken to equal the known value $\alpha_y$; in DGM 1-3, $\Delta^*(\bx_{i})$ is not known but (for the purposes of the simulation study) is estimated by fitting the selection model to a data set of size 1,000,000 before data deletion.
The parameters $\balpha$ cannot be estimated by standard methods, since some $y_i$ are missing, so we use a weighted estimating equation which does not involve the missing $y_i$'s \citep{Rotnitzky++98,Dufouil++04,CNSTAT10}:
$$\sum_i \bx_{Pi} \left\{\frac{r_i}{h(\balpha^T \bx_{Pi} + \Delta^*(\bx_{i}) y_i)}-1\right\} = 0.$$
The substantive model is then fitted to the complete cases with stabilised weights $\hat{p}(r_i=1|\bx_{Si})/\hat{p}(r_i=1|y_i,\bx_{i})$, where $\hat{p}(r_i=1|\bx_{Si})$ is estimated by the same procedure as $\hat{p}(r_i=1|y_i,\bx_{i})$ but with no $\Delta^*(\bx_{i}) y_i$ or $\bx_{Ai}$ terms \citep{Robins++00}.
Variances are computed by the sandwich variance formula, ignoring uncertainty in $\hat{\balpha}$.

\subsection{Estimand}

The estimand of interest is the coefficient $\beta_{Sz}$ in the substantive model. It is computed by fitting the substantive model to the data set of size 1,000,000 before data deletion.
We explore bias, empirical and model-based standard errors, and coverage of estimates $\hat{\beta}_{Sz}$.

\subsection{Results}

Results are shown in Table \ref{tab:simout}.
CC is always biased, often inefficient, and poorly covering.
Small bias (at most 3\% of the true value) is observed in the ``Full'' analysis (i.e.\ before data deletion) in some settings: this is a small-sample effect \citep{Nemes++09}, since as noted above, the true value and ``Full'' are calculated in the same way with large and small samples respectively.
Taking ``Full'' as a gold standard, MS, MI and SM methods all have minimal bias (at most 2\% of true values).
Precisions of MS and MI are similar, with SM slightly inferior in DGM 2.
Coverages are near 95\%, with some over-coverage for SM in some settings, as a consequence of slightly overestimated standard errors (results not shown) due to ignoring uncertainty in $\hat{\balpha}$ \citep{LuncefordDavidian04}.
The performance of MS is not appreciably worse when the selection model is mis-specified (DGM 3) or when the pattern-mixture model is mis-specified (DGM 4).
Computation times for MI are 15-18 times longer than for MS, which is 10-30\% longer than SM.


\section{Discussion} \label{sec:discuss}


We have proposed a mean score method which works well when the sensitivity parameters are known. In practice, of course, the sensitivity parameters are unknown, and a range of values will be used in a sensitivity analysis.

The main practical difficulty in implementing any principled sensitivity analysis is choosing the value(s) of the sensitivity parameters. This is a subjective process requiring subject-matter knowledge and is best done by discussion between statisticians and suitable `experts', typically the trial investigators.
By using the pattern-mixture model, we use a sensitivity parameter $\Delta(\bx_{i})$ that is easier to communicate with `experts' than the corresponding parameters in selection models or shared parameter models.
The procedure has been successfully applied in several trials 
\citep{ian:miss_elicit}.
Special attention is needed to the possibility that $\Delta(\bx_{i})$ varies between randomised groups, because estimated treatment effects are highly sensitive to such variation \citep{ian:IMpriors}.
As with all aspects of trial analysis, plausible ranges of the $\Delta(\bx_{i})$ parameters should be defined before the data are collected or before any analysis.
An alternative approach would report the ``tipping point'', the value of $\Delta(\bx_{i})$ for which the main results are substantively affected, leaving the reader to make the subjective decisions about the plausibility of more extreme values \citep{LiublinskaRubin14}. Presenting this information could be complex without subjective decisions about the difference in $\Delta(\bx_{i})$ between randomised groups.
Effective methods are therefore needed for eliciting sensitivity parameters.

Our method does not incorporate data on discontinuation of treatment, unless this can be included as an auxiliary variable. Our method would however be a suitable adjunct to estimation of effectiveness in a trial with good follow-up after discontinuation of treatment. Further work is needed to combine our sensitivity analysis with models for outcome before and after discontinuation of treatment \citep{LittleYau96}.
In a drug trial in which follow-up ends on discontinuation of treatment, we see our method as estimating efficacy or a \emph{de jure} estimand; if effectiveness or a \emph{de facto} estimand is required then post-discontinuation missing data in each arm may be imputed by the methods of \citet{Carpenter++13}. Further work is needed to perform a full sensitivity analysis in this setting.

Our use of parametric models makes our results susceptible to model mis-specification, and indeed in many cases models (\ref{eq:SM}) and  (\ref{eq:PM}) cannot both be correctly specified except under MAR. However, the simulation study shows that the impact of such model inconsistency is small relative to the impact of assumptions about the missing data and the difficulty of knowing the values of the sensitivity parameters.

We compared the mean score method with multiple imputation and inverse probability weighting. In the case of a quantitative outcome, MI can be simplified by imputing under MAR and then adding the offset to the imputed data before fitting the substantive model and applying Rubin's rules.
We could also impute under MAR and then use a weighted version of Rubin's rules to allow for MNAR \citep{Carpenter++07}. Both MI methods are subject to Monte Carlo error and so seem inferior to the mean score method.
A full likelihood-based analysis of the selection model would also be possible, and a Bayesian analysis could directly allow for uncertainty about $\Delta(\bx_{i})$ in a single analysis \citep{MasonA++12}. These alternative approaches are both more computationally complex.


The proposed mean score method can be extended in various ways.
We have illustrated the method for departures from a MAR assumption, but it can equally be used if the primary analysis with a binary outcome assumed ``missing = failure'', by varying $\Delta(\bx_{i})$ from $-\infty$ rather than from 0.
The method is also appropriate in observational studies, except that mean imputation for missing covariates is not appropriate in this context.
The method can be applied to a cluster-randomised trial as described in Section \ref{app:cluster} of the Supplementary Materials.
Further work could allow the imputation and substantive models to have different link functions, non-canonical links to be used with suitable modification to $\bU_S$, and extension to trials with more than two arms.


Sensitivity analysis should be more widely used to assess the importance of departures from assumptions about missing data.
The proposed mean score approach provides data analysts with a fast and fully theoretically justified way to perform the sensitivity analyses.
It is implemented in a Stata module \texttt{rctmiss} available from Statistical Software Components (SSC) at \url{https://ideas.repec.org/s/boc/bocode.html}.

\section*{Supplementary materials}

The supplementary materials give details of the sandwich variance in equation (\ref{eq:sandwich}) and sketch an extension to clustered data.


\section*{Acknowledgements}
IRW was funded by the Medical Research Council [Unit Programme number U105260558].
NJH was funded by NIH grant 5R01MH054693-12.
JC was funded by ESRC Research Fellowship RES-063-27-0257.
We thank the QUATRO Trial Team for access to the study data.


\bibliographystyle{\home/latex/bst/jrss}
\bibliography{\home/refs/library}

\begin{thebibliography}{47}

\bibitem[Carpenter \emph{et~al.}(2007)Carpenter, Kenward and
  White]{Carpenter++07}
Carpenter, J.~R., Kenward, M.~G. and White, I.~R. (2007) {Sensitivity analysis
  after multiple imputation under missing at random: a weighting approach}.
\newblock \emph{Statistical Methods in Medical Research}, \textbf{16}(3),
  259--275.

\bibitem[Carpenter \emph{et~al.}(2013)Carpenter, Roger and
  Kenward]{Carpenter++13}
Carpenter, J.~R., Roger, J.~H. and Kenward, M.~G. (2013) {Analysis of
  longitudinal trials with protocol deviation: a framework for relevant,
  accessible assumptions, and inference via multiple imputation}.
\newblock \emph{Journal of Biopharmaceutical Statistics}, \textbf{23}(3),
  1352--71.

\bibitem[Daniels and Hogan(2000)]{DanielsHogan00}
Daniels, M.~J. and Hogan, J.~W. (2000) {Reparameterizing the Pattern Mixture
  Model for Sensitivity Analyses Under Informative Dropout}.
\newblock \emph{Biometrics}, \textbf{56}(4), 1241--1248.

\bibitem[Diggle and Kenward(1994)]{DiggleKenward94}
Diggle, P. and Kenward, M.~G. (1994) {Informative drop-out in longitudinal data
  analysis}.
\newblock \emph{Applied Statistics}, \textbf{43}(1), 49--93.

\bibitem[Dufouil \emph{et~al.}(2004)Dufouil, Brayne and Clayton]{Dufouil++04}
Dufouil, C., Brayne, C. and Clayton, D. (2004) {Analysis of longitudinal
  studies with death and drop-out: a case study}.
\newblock \emph{Statistics in Medicine}, \textbf{23}(14), 2215--2226.

\bibitem[Follmann and Wu(1995)]{FollmannWu95}
Follmann, D. and Wu, M. (1995) {An approximate generalized linear model with
  random effects for informative missing data}.
\newblock \emph{Biometrics}, \textbf{51}, 151--168.

\bibitem[Gray \emph{et~al.}(2006)Gray, Leese, Bindman, Becker, Burti, David,
  Gournay, Kikkert, Koeter, Puschner, Schene, Thornicroft and
  Tansella]{Gray++06}
Gray, R., Leese, M., Bindman, J., Becker, T., Burti, L., David, A., Gournay,
  K., Kikkert, M., Koeter, M., Puschner, B., Schene, A., Thornicroft, G. and
  Tansella, M. (2006) {Adherence therapy for people with schizophrenia:
  European multicentre randomised controlled trial}.
\newblock \emph{The British Journal of Psychiatry}, \textbf{189}(6), 508--514.

\bibitem[Groenwold \emph{et~al.}(2012)Groenwold, White, Donders, Carpenter,
  Altman and Moons]{ian:MissInd}
Groenwold, R. H.~H., White, I.~R., Donders, A. R.~T., Carpenter, J.~R., Altman,
  D.~G. and Moons, K. G.~M. (2012) {Missing covariate data in clinical
  research: when and when not to use the missing-indicator method for
  analysis}.
\newblock \emph{Canadian Medical Association Journal}, \textbf{184}(11),
  1265--1269.

\bibitem[Hedeker and Gibbons(2006)]{HedekerGibbons06}
Hedeker, D.~R. and Gibbons, R.~D. (2006) \emph{{Longitudinal data analysis}}.
\newblock John Wiley and Sons.

\bibitem[Higgins \emph{et~al.}(2008)Higgins, White and Wood]{ian:metamiss0}
Higgins, J.~P., White, I.~R. and Wood, A.~M. (2008) {Imputation methods for
  missing outcome data in meta-analysis of clinical trials}.
\newblock \emph{Clinical Trials}, \textbf{5}(3), 225--239.

\bibitem[Jackson \emph{et~al.}(2010)Jackson, White and Leese]{ian:QUATRO_MNAR}
Jackson, D., White, I.~R. and Leese, M. (2010) {How much can we learn about
  missing data?: an exploration of a clinical trial in psychiatry}.
\newblock \emph{Journal of the Royal Statistical Society: Series A (Statistics
  in Society)}, \textbf{173}(3), 593--612.

\bibitem[Kaciroti \emph{et~al.}(2009)Kaciroti, Schork, Raghunathan and
  Julius]{Kaciroti++09}
Kaciroti, N.~A., Schork, M.~A., Raghunathan, T. and Julius, S. (2009) {A
  Bayesian sensitivity model for intention-to-treat analysis on binary outcomes
  with dropouts}.
\newblock \emph{Statistics in Medicine}, \textbf{28}(4), 572--585.

\bibitem[Kauermann and Carroll(2001)]{KauermannCarroll01}
Kauermann, G. and Carroll, R.~J. (2001) {A note on the efficiency of sandwich
  covariance matrix estimation}.
\newblock \emph{Journal of the American Statistical Association},
  \textbf{96}(456), 1387--1396.

\bibitem[Kenward \emph{et~al.}(2001)Kenward, Goetghebeur and
  Molenberghs]{Kenward++01}
Kenward, M., Goetghebeur, E. and Molenberghs, G. (2001) {Sensitivity analysis
  for incomplete categorical data}.
\newblock \emph{Statistical Modelling}, \textbf{1}(1), 31--48.

\bibitem[Kenward(1998)]{Kenward98}
Kenward, M.~G. (1998) {Selection models for repeated measurements with
  non-random dropout: an illustration of sensitivity}.
\newblock \emph{Statistics in Medicine}, \textbf{17}(23), 2723--2732.

\bibitem[Little and Yau(1996)]{LittleYau96}
Little, R. and Yau, L. (1996) {Intent-to-treat analysis for longitudinal
  studies with drop-outs}.
\newblock \emph{Biometrics}, \textbf{52}(4), 1324--1333.

\bibitem[Little(1993)]{Little93}
Little, R. J.~A. (1993) {Pattern-mixture models for multivariate incomplete
  data}.
\newblock \emph{Journal of the American Statistical Association},
  \textbf{88}(421), 125--134.

\bibitem[Little(1994)]{Little94}
Little, R. J.~A. (1994) {A class of pattern-mixture models for normal
  incomplete data}.
\newblock \emph{Biometrika}, \textbf{81}(3), 471--483.

\bibitem[Little and Rubin(2002)]{LittleRubin02}
Little, R. J.~A. and Rubin, D.~B. (2002) \emph{{Statistical Analysis with
  Missing Data}}.
\newblock Hoboken, N. J.: Wiley, second edition.

\bibitem[Liublinska and Rubin(2014)]{LiublinskaRubin14}
Liublinska, V. and Rubin, D.~B. (2014) {Sensitivity analysis for a partially
  missing binary outcome in a two-arm randomized clinical trial}.
\newblock \emph{Statistics in Medicine}, \textbf{33}(24), 4170--4185.

\bibitem[Lu \emph{et~al.}(2007)Lu, Preisser, Qaqish, Suchindran, Bangdiwala and
  Wolfson]{Lu++07}
Lu, B., Preisser, J.~S., Qaqish, B.~F., Suchindran, C., Bangdiwala, S.~I. and
  Wolfson, M. (2007) {A comparison of two bias-corrected covariance estimators
  for generalized estimating equations}.
\newblock \emph{Biometrics}, \textbf{63}(3), 935--941.

\bibitem[Lunceford and Davidian(2004)]{LuncefordDavidian04}
Lunceford, J.~K. and Davidian, M. (2004) {Stratification and weighting via the
  propensity score in estimation of causal treatment effects: a comparative
  study}.
\newblock \emph{Statistics in Medicine}, \textbf{23}, 2937--2960.

\bibitem[Mason \emph{et~al.}(2012)Mason, Richardson, Plewis and
  Best]{MasonA++12}
Mason, A., Richardson, S., Plewis, I. and Best, N. (2012) {Strategy for
  modelling nonrandom missing data mechanisms in observational studies using
  Bayesian methods}.
\newblock \emph{Journal of Official Statistics}, \textbf{28}(2), 279--302.

\bibitem[Mavridis \emph{et~al.}(2015)Mavridis, White, Higgins, Cipriani and
  Salanti]{ian:metamisscts}
Mavridis, D., White, I.~R., Higgins, J. P.~T., Cipriani, A. and Salanti, G.
  (2015) {Allowing for uncertainty due to missing continuous outcome data in
  pair-wise and network meta-analysis}.
\newblock \emph{Statistics in Medicine}, \textbf{34}, 721--741.

\bibitem[{National Research Council}(2010)]{CNSTAT10}
{National Research Council} (2010) \emph{{The Prevention and Treatment of
  Missing Data in Clinical Trials}}.
\newblock Washington, DC: Panel on Handling Missing Data in Clinical Trials.
  Committee on National Statistics, Division of Behavioral and Social Sciences
  and Education. The National Academies Press.

\bibitem[Nemes \emph{et~al.}(2009)Nemes, Jonasson, Genell and
  Steineck]{Nemes++09}
Nemes, S., Jonasson, J.~M., Genell, A. and Steineck, G. (2009) {Bias in odds
  ratios by logistic regression modelling and sample size}.
\newblock \emph{BMC Medical Research Methodology}, \textbf{9}(1), 56.

\bibitem[Pepe \emph{et~al.}(1994)Pepe, Reilly and Fleming]{Pepe++94}
Pepe, M.~S., Reilly, M. and Fleming, T.~R. (1994) {Auxiliary outcome data and
  the mean score method}.
\newblock \emph{Journal of Statistical Planning and Inference}, \textbf{42},
  137--160.

\bibitem[Reilly and Pepe(1995)]{ReillyPepe95}
Reilly, M. and Pepe, M.~S. (1995) {A mean score method for missing and
  auxiliary covariate data in regression models}.
\newblock \emph{Biometrika}, \textbf{82}(2), 299--314.

\bibitem[Robins \emph{et~al.}(2000)Robins, Hernan and Brumback]{Robins++00}
Robins, J.~M., Hernan, M.~A. and Brumback, B. (2000) {Marginal structural
  models and causal inference in epidemiology}.
\newblock \emph{Epidemiology}, \textbf{11}, 550--560.

\bibitem[Rogers(1993)]{Rogers93*1}
Rogers, W.~H. (1993) {Regression standard errors in clustered samples}.
\newblock \emph{Stata Technical Bulletin}, \textbf{13}, 19--23.

\bibitem[Rotnitzky \emph{et~al.}(1998)Rotnitzky, Robins and
  Scharfstein]{Rotnitzky++98}
Rotnitzky, A., Robins, J.~M. and Scharfstein, D.~O. (1998) {Semiparametric
  regression for repeated outcomes with nonignorable nonresponse}.
\newblock \emph{Journal of the American Statistical Association},
  \textbf{93}(444), 1321--1339.

\bibitem[Roy(2003)]{Roy03}
Roy, J. (2003) {Modeling longitudinal data with nonignorable dropouts using a
  latent dropout class model}.
\newblock \emph{Biometrics}, \textbf{59}(4), 829--836.

\bibitem[Rubin(1987)]{Rubin87}
Rubin, D.~B. (1987) \emph{{Multiple Imputation for Nonresponse in Surveys}}.
\newblock New York: John Wiley and Sons.

\bibitem[Scharfstein \emph{et~al.}(2014)Scharfstein, McDermott, Olson and
  Wiegand]{Scharfstein++14}
Scharfstein, D., McDermott, A., Olson, W. and Wiegand, F. (2014) {Global
  sensitivity analysis for repeated measures studies with informative dropout:
  a fully parametric approach}.
\newblock \emph{Statistics in Biopharmaceutical Research}, \textbf{6}(4),
  338--348.

\bibitem[Scharfstein \emph{et~al.}(2003)Scharfstein, Daniels and
  Robins]{Scharfstein++03a}
Scharfstein, D.~O., Daniels, M.~J. and Robins, J.~M. (2003) {Incorporating
  prior beliefs about selection bias into the analysis of randomized trials
  with missing outcomes}.
\newblock \emph{Biostatistics}, \textbf{4}, 495--512.

\bibitem[StataCorp(2011)]{Stata12}
StataCorp (2011) \emph{{Stata Statistical Software: Release 12}}.
\newblock College Station, TX: StataCorp LP.

\bibitem[Ware(1993)]{Ware93}
Ware, J.~E. (1993) \emph{{SF-36 Health Survey: Manual and Interpretation
  Guide}}.
\newblock Boston: The Health Institute, New England Medical Center.

\bibitem[Ware(2003)]{Ware03}
Ware, J.~H. (2003) {Interpreting incomplete data in studies of diet and weight
  loss}.
\newblock \emph{New England Journal of Medicine}, \textbf{348}(21), 2136--2137.

\bibitem[West \emph{et~al.}(2005)West, Hajek, Stead and Stapleton]{West++05}
West, R., Hajek, P., Stead, L. and Stapleton, J. (2005) {Outcome criteria in
  smoking cessation trials: proposal for a common standard}.
\newblock \emph{Addiction}, \textbf{100}, 299--303.

\bibitem[White(2015)]{ian:miss_elicit}
White, I.~R. (2015) {Sensitivity analysis: The elicitation and use of expert
  opinion}.
\newblock In \emph{Handbook of Missing Data Methodology} (Eds G.~Molenberghs,
  G.~Fitzmaurice, M.~G. Kenward, A.~Tsiatis and G.~Verbeke). Chapman and Hall.

\bibitem[White and Thompson(2005)]{ian:MSGBSL}
White, I.~R. and Thompson, S.~G. (2005) {Adjusting for partially missing
  baseline measurements in randomized trials}.
\newblock \emph{Statistics in Medicine}, \textbf{24}(7), 993--1007.

\bibitem[White \emph{et~al.}(2007)White, Carpenter, Evans and
  Schroter]{ian:IMpriors}
White, I.~R., Carpenter, J., Evans, S. and Schroter, S. (2007) {Eliciting and
  using expert opinions about dropout bias in randomized controlled trials}.
\newblock \emph{Clinical Trials}, \textbf{4}(2), 125--139.

\bibitem[White \emph{et~al.}(2008)White, Higgins and Wood]{ian:metamiss1}
White, I.~R., Higgins, J. P.~T. and Wood, A.~M. (2008) {Allowing for
  uncertainty due to missing data in meta-analysis - Part 1: Two-stage
  methods}.
\newblock \emph{Statistics in Medicine}, \textbf{27}(5), 711--727.

\bibitem[White \emph{et~al.}(2011{a})White, Horton, Carpenter and
  Pocock]{ian:ITTmed}
White, I.~R., Horton, N.~J., Carpenter, J. and Pocock, S.~J. (2011{a})
  {Strategy for intention to treat analysis in randomised trials with missing
  outcome data}.
\newblock \emph{British Medical Journal}, \textbf{342}, d40.

\bibitem[White \emph{et~al.}(2011{b})White, Wood and Royston]{ian:MItutorial}
White, I.~R., Wood, A. and Royston, P. (2011{b}) {Tutorial in biostatistics:
  Multiple imputation using chained equations: issues and guidance for
  practice}.
\newblock \emph{Stat Med}, \textbf{30}, 377--399.

\bibitem[White \emph{et~al.}(2012)White, Carpenter and Horton]{ian:ITTstat}
White, I.~R., Carpenter, J. and Horton, N.~J. (2012) {Including all individuals
  is not enough: lessons for intention-to-treat analysis}.
\newblock \emph{Clinical Trials}, \textbf{9}, 396--407.

\bibitem[Wood \emph{et~al.}(2004)Wood, White and Thompson]{ian:missing_survey}
Wood, A.~M., White, I.~R. and Thompson, S.~G. (2004) {Are missing outcome data
  adequately handled? A review of published randomized controlled trials in
  major medical journals}.
\newblock \emph{Clinical Trials}, \textbf{1}(4), 368--376.

\end{thebibliography}


\clearpage
\renewcommand{\baselinestretch}{1}
\begin{table}[t]
\caption{QUATRO trial: data summary.}
\label{tab:data}
\centering
\vspace{1ex}
\begin{tabular}{llcc}\hline
&                            &  Intervention  & Control       \\
&                            &  (n=204)       & (n=205)       \\\hline
Centre&    Amsterdam (\%)    &    50 (25\%)   &   50  (24\%)  \\
       &   Leipzig   (\%)    &    49 (24\%)   &   48  (23\%)  \\
        &  London    (\%)    &    45 (22\%)   &   47  (23\%)  \\
         & Verona    (\%)    &    60 (29\%)   &   60  (29\%)  \\\hline
MCS at baseline & Mean (SD)  &  38.4 (11.2)   &  40.1 (12.1)  \\
        & Missing (\%)       &    13  (6\%)   &  10    (5\%)  \\\hline
MCS at 1 year   & Mean (SD)  &  40.2 (12.0)   &  41.3 (11.5)  \\
            & $>40$ (\%)     &  99   (57\%)   &  104  (54\%)  \\
        & Missing (\%)       &  29   (14\%)   &  13    (6\%)  \\\hline
\end{tabular}
\end{table}


\begin{table}[t]
\caption{Simulation study: data generating models.}
\label{tab:simdesc}
\centering
\vspace{1ex}
\begin{tabular}{lcccc}\hline
DGM & Type            & PM correct? & SM correct? & Auxiliary variables? \\\hline
1   & Pattern-mixture & Yes         & Yes         & No  \\
2   & Pattern-mixture & Yes         & Yes         & Yes \\
3   & Pattern-mixture & Yes         & No          & No  \\
4   & Selection       & No          & Yes         & No  \\\hline
\end{tabular}
\\[1ex]
\begin{tabular}{lc}\hline
Parameters & Description \\\hline
a   & Base case \\
b   & Larger sample size \\
c   & More missing data \\
d   & Larger departure from MAR  \\\hline
\end{tabular}
\end{table}

\begin{landscape}
\addtocounter{page}{-1}
\begin{table}[t]
\caption{Simulation results: bias, empirical standard error and coverage of nominal 95\% confidence interval for methods Full (before data deletion), CC (complete cases analysis), MS (mean score), MI (multiple imputation), SM (selection model + IPW). Error denotes maximum Monte Carlo error.}
\label{tab:simout}
\centering
\vspace{1ex}
\begin{tabular}{l|ccccc|ccccc|ccccc}\hline
DGM&\multicolumn{5}{c|}{Bias}&\multicolumn{5}{c|}{Empirical SE}&\multicolumn{5}{c}{Coverage} \\ \cline{2-16}
& Full & CC & MS & MI & SM & Full & CC & MS & MI & SM & Full & CC & MS & MI & SM \\ \hline
1a&0.007&-0.128&0.010&0.010&0.009&0.191&0.223&0.218&0.219&0.218&95.0&90.6&95.1&95.3&95.8\\
1b&0.004&-0.130&0.007&0.006&0.006&0.095&0.113&0.111&0.112&0.111&94.3&76.7&93.8&93.4&94.6\\
1c&0.007&-0.167&0.018&0.018&0.018&0.183&0.267&0.258&0.259&0.258&96.5&91.9&95.9&95.4&96.6\\
1d&0.000&-0.177&0.003&0.004&0.004&0.187&0.223&0.203&0.204&0.203&95.1&88.3&95.6&95.5&97.1\\
\hline
2a&0.009&-0.151&0.010&0.010&0.005&0.178&0.219&0.203&0.203&0.209&96.3&90.1&95.4&95.4&97.0\\
2b&0.004&-0.157&0.003&0.003&-0.001&0.092&0.109&0.101&0.101&0.104&94.6&71.5&95.1&95.3&96.9\\
2c&0.005&-0.206&0.002&-0.001&-0.004&0.185&0.296&0.256&0.257&0.290&94.9&88.1&95.0&94.7&97.2\\
2d&0.011&-0.175&0.011&0.010&0.007&0.186&0.227&0.200&0.200&0.202&95.0&86.1&94.5&95.2&98.2\\
\hline
3a&0.016&-0.110&0.021&0.020&0.015&0.208&0.249&0.245&0.246&0.249&96.5&92.7&95.4&94.7&95.3\\
3b&0.003&-0.127&0.003&0.003&-0.003&0.106&0.122&0.120&0.120&0.121&94.8&81.8&95.3&95.2&95.3\\
3c&0.012&-0.160&0.015&0.012&0.018&0.217&0.327&0.316&0.318&0.323&95.1&90.3&95.0&95.1&95.4\\
3d&0.019&-0.163&0.020&0.020&0.020&0.216&0.254&0.238&0.239&0.249&94.2&88.6&94.5&94.6&95.3\\
\hline
4a&0.005&-0.148&0.014&0.012&0.006&0.218&0.255&0.251&0.252&0.253&94.2&90.0&94.8&94.5&95.0\\
4b&0.006&-0.147&0.015&0.015&0.008&0.107&0.124&0.122&0.122&0.123&94.0&77.1&94.0&94.1&94.6\\
4c&0.008&-0.203&0.004&0.002&0.011&0.213&0.345&0.336&0.337&0.343&94.2&90.8&94.2&94.3&95.0\\
4d&0.026&-0.118&0.045&0.045&0.036&0.211&0.250&0.246&0.248&0.249&94.2&91.2&94.3&94.5&94.8\\
\hline
Error&0.007&0.011&0.011&0.011&0.011&0.005&0.008&0.008&0.008&0.008&0.8&1.4&0.8&0.8&0.7\\
\hline
\end{tabular}

\end{table}														
\end{landscape}

\addtocounter{page}{-1}


\vfill


\clearpage
\clearpage
\setcounter{page}{1}
\appendix
\section*{Supplementary materials}

\section{Details of mean score variance}\label{app:variance}

The sandwich variance (\ref{eq:sandwich}) is computed from $\bB$ and $\bC$, where
$\bB$ has components
\begin{eqnarray*}
\bB_{SS} &= -d\bU_S/d\bbeta_S &=  \sum_i h'(\hat{\bbeta}_S \bx_{Si}) \bx_{Si} \bx_{Si}^T \\
\bB_{SP} &= -d\bU_S/d\bbeta_P &= -\sum_i (1-r_i) h'(\hat{\bbeta}_P \bx_{Pi} + \Delta_{i}) \bx_{Si} \bx_{Pi}^T \\
\bB_{PS} &= -d\bU_P/d\bbeta_S &=  \bzero \\
\bB_{PP} &= -d\bU_P/d\bbeta_P &=  \sum_i r_i h'(\hat{\bbeta}_P \bx_{Pi}) \bx_{Pi} \bx_{Pi}^T
\end{eqnarray*}
and $\bC$ has components
\begin{eqnarray*}
\bC_{SS} &=& \sum_i e_{Si}^2 \bx_{Si} \bx_{Si}^T \\
\bC_{SP} =\bC_{PS}^T &=& \sum_i e_{Si}e_{Pi} \bx_{Si} \bx_{Pi}^T  \\
\bC_{PP} &=& \sum_i e_{Pi}^2 \bx_{Pi} \bx_{Pi}^T
\end{eqnarray*}
where $e_{Si}=y^*_i(\hat{\bbeta}_P) - h(\hat{\bbeta}_S^T \bx_{Si})$ and $e_{Pi}=r_i\{y_i - h(\hat{\bbeta}_P^T \bx_{Pi})\}$.

\section{Modifications for clustered data}\label{app:cluster}

If data are clustered, as in a cluster-randomised trial, we need to modify the variance calculations in Sections \ref{sec:fullsandwich} and \ref{sec:twolinreg} and the small-sample corrections in Sections \ref{sec:fullsandwich:small} and \ref{sec:twolinreg:small}. 
Let $m$ be the total number of clusters, $m_{obs}$ be the number of clusters with at least one observed outcome, and $m_{mis} = m - m_{obs}$ be the number of clusters with no observed outcome.
Let the data be subscripted by cluster membership $c=1,\ldots,m$ as well as individual $i$.

For the full sandwich variance method of Section \ref{sec:fullsandwich}, we only need to redefine the matrix
$\bC=\sum_c \bU_c(\hat{\bbeta}) \bU_c(\hat{\bbeta})^T$
where $\bU_c(\hat{\bbeta})=\sum_i \bU_{ci}(\hat{\bbeta})$
\citep{Rogers93*1}.

For the two linear regressions method of Section \ref{sec:twolinreg}, we similarly take $\varb{\hat{\bbeta}_P}$ and $\varb{\hat{\bbeta}_S-\hat{\bbeta}_P}$ as clustered sandwich variances.

\newcommand{\meff}{m_{\mbox{\it\scriptsize eff}}}

For the small-sample methods of Section \ref{sec:equiv}, we assume the standard methods use a small-sample correction factor $f=\frac{n-1}{n-\pstar} \frac{m}{m-1}$, and use $m-1$ degrees of freedom for linear regression \citep{Stata12}. We replace $n$ and $m$ by $\neff$ and $\meff$, calculated by the two methods explained below.

For the  full sandwich variance method, we compute $\neff$ as in Section \ref{sec:fullsandwich:small}, and compute $\meff = m_{obs} + (I_{mis}/I_{mis^*}) m_{mis}$.

For the two linear regressions method, 
the variance with small-sample correction is (as before) 
$\varhat{\hat{\bbeta}_P}+ \varhat{\hat{\bbeta}_S-\hat{\bbeta}_P}$ $= V_{small}$.
The corresponding variance without small-sample correction is
$\frac{n_{obs}-p}{n_{obs}-1} \frac{m_{obs}-1}{m_{obs}} \varhat{\hat{\bbeta}_P}
+
\frac{n-p}{n-1} \frac{m-1}{m} \varhat{\hat{\bbeta}_S-\hat{\bbeta}_P}
=
V_{large}$.
The heuristic $V_{small} \approx \frac{\neff-1}{\neff-p} \frac{\meff}{\meff-1} V_{large}$ leads to the equation 
$|V_{small}| = \left(\frac{\neff}{\neff-p}\right)^{p} |V_{large}|$.
However, we have two unknowns $\neff$ and $\meff$, so we take a second equation representing the variance with small-sample correction only for the number of clusters:
$\frac{m_{obs}-1}{m_{obs}} \varhat{\hat{\bbeta}_P}
+
\frac{m-1}{m} \varhat{\hat{\bbeta}_S-\hat{\bbeta}_P}
=
V_{large n}$ say, with the heuristic $V_{small} \approx \frac{\meff}{\meff-1} V_{large n}$ and the second equation
$|V_{small}| = \left(\frac{\meff}{\meff-1}\right)^{p} |V_{large n}|$.
We solve the second equation for $\meff$ and then the first equation for $\neff$.

\end{document}